\documentclass[preprint,nofootinbib]{revtex4}

\usepackage{graphicx}
\begin{document}

\title{Improved Predictions for Higgs $Q_T$ at the Tevatron and the LHC}

\author{Qing-Hong Cao$^{1,2}$, Chuan-Ren Chen$^{3}$, Carl Schmidt$^{4}$, Zhao Li$^{4}$ and C.-P. Yuan$^{4}$}

\affiliation{${}^{1}$HEP Divison, Argonne National Laboratory, Argonne, IL 60439, USA\\
${}^{2}$Enrico Fermi Institute, University of Chicago, Chicago, IL 60637, USA\\
${}^{3}$Institute for the Physics and Mathematics of the Universe,
University of Tokyo, Chiba 277-8568, Japan\\
 ${}^{4}$Department of Physics and Astronomy, Michigan State University, East
Lansing, MI 48824, USA }

\begin{abstract}
The search for the Higgs boson at the Tevatron and the LHC relies on detailed calculations of the kinematics of Higgs boson production and decay.
In this paper, we improve the calculation of the distribution in transverse momentum, $Q_T$, of the Higgs boson in the gluon fusion production process, $gg\to H$,
by matching the resummed distribution at small $Q_T$ with the ${\cal O}(\alpha_{s}^4)$ fixed-order perturbative calculation at high $Q_T$ in the ResBos
Monte Carlo program.  The distribution is higher at large $Q_T$ than with the old ${\cal O}(\alpha_{s}^3)$ fixed-order perturbative calculation, and the
matching with the resummed calculation is much smoother.  The total cross section is also increased, more in line with next-to-next-to-leading-order calculations. 
We also study the effect of the new calculation on the distribution of $\Delta\phi_{\ell\ell}$ in the overall process $gg\to H\to W^{+} W^{-}\to\ell^{+}\ell^{-}\nu\bar{\nu}$, and
the effect of PDF uncertainties on the distributions at the Tevatron and the LHC.
\end{abstract}

\preprint{ANL-HEP-PR-09-20, EFI-09-14, IPMU09-0038, MSUHEP-090911}

\maketitle

\section{Introduction}
\label{sec:intro}
It has been over 25 years since the discovery of the $W$ and $Z$
gauge bosons in the UA1 and UA2 colliders at CERN, and we are finally
on the verge of discovering the source of their mass. With the on-going
studies at the Tevatron collider at Fermilab and the turn-on
of the Large Hadron Collider (LHC) at CERN, we will finally be able
to probe directly the physics that breaks the electroweak symmetry
and distinguishes the massive $W$ and $Z$ bosons from the massless
photon. The simplest model of Electroweak Symmetry Breaking (EWSB)
is the Standard Model (SM) of particle physics, which contains
a complex electroweak scalar doublet that acquires a vacuum expectation
value, thereby breaking the electroweak symmetry spontaneously. Three
of the degrees of freedom of this complex doublet become the longitudinal
modes of the massive $W^{+}$, $W^{-}$, and $Z$ bosons, while the
remaining degree of freedom is manifested as a single neutral scalar---the
SM Higgs boson ($H$). Although this is not the only possible mechanism
for EWSB, the search for the Higgs boson is the benchmark study for
EWSB physics to be undertaken at the Tevatron and the LHC. If all
goes well, it should be observed or ruled out within the next few
years.

The most stringent limits on the Higgs boson mass, $m_{H}$, come
from direct searches for the particle at LEP2 in the process $e^{+}e^{-}\rightarrow ZH$,
where the lower bound of 114.1 GeV has been obtained at 95\% confidence
level (C.L.)~\cite{Barate:2003sz}.  In addition, preliminary results from a combined fit of
CDF and D\O~data at the Tevatron has been used to exclude the mass range of
$160\,\rm{GeV} < m_{H} < 170\, \rm{GeV}$ at 95\% C.L.~\cite{Phenomena:2009pt}~\
Beyond the direct search for a real Higgs boson, the effect of virtual Higgs
bosons in loop calculations can be used to obtain indirect bounds
on $m_{H}$. Current global fits to electroweak precision measurements,
in combination with the direct search limit, prefer {$m_{H}\lesssim191$}
GeV at 95\% confidence level~\cite{lepewwg}. The Tevatron collider has a reasonable chance of discovery or exclusion over much of this
preferred range of $m_{H}$, assuming 7 fb$^{-1}$ of data obtained
by the end of Tevatron running~\cite{Duperrin:2008in}. Furthermore,
the LHC can be expected to cover the entire range of Higgs boson masses
up to about $1\,\rm{TeV}$, which is a rough upper bound on $m_{H}$, based
on triviality and unitarity of the Standard Model~\cite{Kolda:2000wi}.

At both the Tevatron and the LHC the largest channel for production
of the Higgs boson is gluon fusion, with the $ggH$ coupling arising
via (mainly, top and bottom) quark loops. Other important channels are production of the
Higgs boson with an associated $W$ boson, $Z$ boson, or top quark
pairs, as well as production of the Higgs boson through vector-boson
or bottom-quark fusion. The importance of each channel for discovery/exclusion
of the Higgs boson depends on its mass. A light Higgs boson ($m_{H}\lesssim135$
GeV) decays predominantly to bottom quark pairs. In this case the
inclusive Higgs boson signal is very difficult to pick out from the
very large QCD $b\bar{b}$ background. The best sensitivity at the
Tevatron in this mass range is instead found in the $WH$ and $ZH$
associated production channels, where the extra particles can be used
to better distinguish the signal from background. A heavier Higgs
boson ($m_{H}\gtrsim135$ GeV) decays predominantly to $W$ boson
pairs with one of the $W$'s potentially off-shell. In this mass range
the inclusive production through gluon fusion is most important at
the Tevatron, with the best sensitivity occurring around $m_{H}\approx160-170$
GeV, where the $WW$ decay mode is fully open. These
modes and many others were used in the recent combined fit at the Tevatron
to exclude  the mass range of
$160\,{\rm GeV} < m_{H} < 170\, \rm{GeV}$ at 95\% C.L.~\cite{Phenomena:2009pt}~\ Other important
Higgs decay modes, both at the Tevatron and the LHC, are $ZZ$ for
high mass Higgs bosons, and tau pairs and photon pairs for low mass
Higgs bosons.

In order to best discern the Higgs boson signal from background, it
is necessary to have the most accurate predictions possible for the
kinematic distributions of the Higgs boson. In the leading order (LO) calculation
of the $gg\rightarrow H+X$ cross section, the Higgs boson is produced
with exactly zero transverse momentum, $Q_{T}=0$. In higher order
calculations it can have non-zero $Q_{T}$, due to the emission of
additional gluons or quarks, but the calculation at any fixed order
of perturbation theory diverges as $Q_{T}\rightarrow0$. Thus, any
fixed-order calculation is unsuitable for the study of the $Q_{T}$-dependence
of the Higgs boson (except at large $Q_{T}$), or for the study of any other kinematic
distribution that is strongly affected by soft gluon radiation. Fortunately,
the soft-gluon effects that occur for small $Q_{T}$ can be incorporated
into the calculation, either by their direct production in a parton
shower Monte Carlo, such as PHYTHIA~\cite{Sjostrand:2007gs} or HERWIG~\cite{Corcella:2002jc},
or by analytic resummation of the associated large logarithms, as
proposed by Collins, Soper, and Sterman (CSS)~\cite{Collins:1981uk,Collins:1981va,Collins:1984kg}.
This systematic resummation in powers of the strong coupling $\alpha_{s}$
times powers of the large logarithm $\ln(Q/Q_{T})$ has been applied
to the Higgs boson process, as well as other processes, in the general
resummation code ResBos~\cite{Balazs:1997xd}. For the present process,
the scale of the resummation, $Q$, is equal to the invariant mass
of the produced Higgs boson, unless otherwise specified.

The resummation of large logarithms at small $Q_{T}$ has been analyzed
for Higgs production in a number of studies in recent years~\cite{Balazs:2000wv,Berger:2002ut,Kulesza:2003wn,Bozzi:2003jy,Bozzi:2005wk,Cao:2007du,Bozzi:2007pn}.
The power of the logarithms that are resummed is determined by parameters,
which can be extracted order-by-order in $\alpha_{s}$ from the perturbative
Higgs production cross sections. The calculation of the Higgs boson
production cross section in gluon-gluon scattering has been calculated
at leading order, next-to-leading order (NLO)~\cite{Dawson:1990zj}
and next-to-next-to-leading order (NNLO)~\cite{Harlander:2002wh,Anastasiou:2002yz,Ravindran:2003um}
in the infinite-top-quark-mass limit, and at LO and NLO~\cite{Djouadi:1991tka,Graudenz:1992pv}
with full top quark mass dependence. In addition to the QCD corrections, the NLO electroweak (EW)
corrections have also been considered in the infinite-top-quark-mass limit~\cite{Djouadi:1994ge},
and more complete calculations have been performed by including light quark
and top quark effects~\cite{Aglietti:2004nj,Degrassi:2004mx}. Recently, the effects of the combined QCD and EW
corrections were analyzed~\cite{Actis:2008ug,Anastasiou:2008tj}.
The inclusive differential cross
section at non-zero $Q_{T}$, which begins at one higher power of
$\alpha_{s}$, has been calculated at NLO in the infinite-top-mass
limit~\cite{deFlorian:1999zd,Ravindran:2002dc,Glosser:2002gm}, and
at LO with full top quark mass dependence~\cite{Ellis:1987xu,Baur:1989cm}.
In the infinite-top-quark-mass limit, the heavy top quark loop contracts
to an effective gluon-Higgs operator, which simplifies the calculation
greatly, effectively reducing the number of loops by one. In addition,
it has been shown, at least at NLO, that it is a good approximation
to calculate in the infinite top quark mass effective theory, while
rescaling by the LO cross section with full top and bottom quark mass dependence~\cite{Kramer:1996iq}.
Thus, it has become standard to use this approximation to compute
the Higgs boson cross section. For nonzero $Q_{T}$ this approximation
is also good as long as $Q_{T}\lesssim m_{t}$ and $m_{H}\lesssim m_{t}$~\cite{Ellis:1987xu,Baur:1989cm}.

A recent analysis using the resummation program ResBos to study the phenomenology
of Higgs boson production at the Tevatron and the LHC was presented
in Ref.~\cite{Cao:2007du}. In that work the CSS resummation at small
$Q_{T}$ was matched onto the LO calculation at large $Q_{T}$. In
this work we have updated the program so that it matches on to the
NLO calculation at large $Q_{T}$, using the code developed in Ref.~\cite{Glosser:2002gm}.
We shall see that this is more consistent with the precision currently
included at small $Q_{T}$ in the resummation program.
The calculation is used to model the Higgs boson $Q_T$ spectrum to get a better 
theoretical prediction at the Tevatron~\cite{Peters:2010tv}. 

The remainder of the paper is organized as follows. In section~\ref{sec:resum}
we give a brief description of the resummation procedure that is implemented
in ResBos. We explain the order of the coefficients used in the resummation
calculation and how they are matched on to the fixed-order calculation at
large $Q_{T}$, and how  the improvement is performed to include ${\cal O} (\alpha_s^4)$. We also calculate the total production cross sections of the Higgs boson at the Tevatron, the LHC with 7 TeV,
10 TeV and 14 TeV center-of-mass (c.m.) energy and compare with NLO results.   In section~\ref{sec:numerical}  we  use the updated code of ResBos to produce various kinematic distributions for the Higgs boson.  In particular, we emphasize the changes
coming from the updated calculation, and note the improvement in the
matching between the low and high $Q_{T}$ regimes. We also discuss
the implications of the new predictions
on the $\Delta \phi_{\ell\ell}$ correlation of the two charged
leptons in the decay $H\rightarrow W^{+}W^{-}\rightarrow\ell^{+}\nu\ell^{-}\bar{\nu}$.
For completeness, we also discuss in section~\ref{sec:scale_dep} the dependence of the resummation prediction
on various choices of the renormalization and factorization scales.
Finally, in section~\ref{sec:conclusion} we give our conclusions.

\section{Soft gluon resummation}
\label{sec:resum}
\subsection{Formalism}
\label{sec:css}
In order to make transparent what we have implemented in the ResBos program, we begin by briefly
reviewing the CSS formalism of soft gluon resummation.   Resumming the soft gluons and using
the narrow width of the Higgs boson to factorize the Higgs production from its subsequent decay,
we can write the inclusive differential cross section for $gg\to H\to VV\to4$ fermions as
\begin{eqnarray}
 \label{eq:resum_formalism}
 &  & \frac{d\sigma(gg\to HX\to VVX\to f_{1}f_{2}f_{3}f_{4}X)}{dQ^{2}dQ_{T}^{2}dyd\phi_{H}d\Pi_{4}}\nonumber \\
 & = &\kappa\, \sigma_{0}\,\frac{Q^{2}}{2S}\,\frac{1}{(Q^{2}-m_{H}^{2})^{2}+(Q^{2}\Gamma_{H}/m_{H})^{2}}
\Biggl|\mathcal{M}(H\to VV\to f_{1}f_{2}f_{3}f_{4})\Biggr|^{2}\\
 & \times & \Biggl\{\frac{1}{(2\pi)^{2}}\int d^{2}b\, e^{iQ_{T}\cdot b}\tilde{W}_{gg}(b_{*},Q,x_{1},x_{2},C_{1,2,3})\tilde{W}_{gg}^{NP}(b,Q,x_{1},x_{2})+Y(Q_{T},Q,x_{1},x_{2},C_{4})\Biggr\}\,,\nonumber
 \end{eqnarray}
where
$S$ is the square of the center-of-mass energy; $V$ and $f_{i}$ denote vector boson and fermion, respectively;
$Q$, $Q_{T}$, $y$, $\phi_{H}$ and $\Gamma_H$ are the invariant mass, transverse
momentum, rapidity, azimuthal angle and total decay width of the Higgs boson, respectively,
defined in the lab frame; and d$\Pi_{4}$ represents the four-body
phase space of the Higgs boson decay, defined in the Collin-Soper
frame~\cite{Collins:1977iv}.
In Eq.~(\ref{eq:resum_formalism}), the quantity
 \begin{equation}
   \label{eq:sigma0}
 \sigma_0\ =\ \frac{\sqrt{2}\,G_F\,\alpha_s^2}{576\,\pi}\,,
 \end{equation}
arises as an overall factor in the infinite-top-quark-mass limit,
where $\alpha_s$ is evaluated at the hard scale $C_2 Q$ with $C_2=1$ being the canonical value.
Furthermore, we have multiplied $\sigma_0$ by
an additional factor
  \begin{equation}
  \label{eq:kappa}
  \kappa\ =\ \frac{\sigma^{\rm LO}(m_t,m_b,m_c)}{\sigma^{\rm LO}(\infty,0,0)}\,,
   \end{equation}
which takes into account the masses of the top, bottom, and charm quarks at LO.  It has been shown that multiplying the NLO Higgs cross section in the infinite-top-quark-mass limit
by the factor $\kappa$ is a good approximation to the full mass-dependent NLO cross section over a wide range of
Higgs boson masses~\cite{Kramer:1996iq}.
In Eq.~(\ref{eq:resum_formalism}), $\left|\mathcal{M}(\cdots)\right|^{2}$ denotes the matrix element
squared of the Higgs boson decay whose analytical expressions are
given in Ref.~\cite{Cao:2007du}.

In Eq.~(\ref{eq:resum_formalism}), the term containing $\tilde{W}_{gg}$ dominates at small $Q_T$, growing as $Q_{T}^{-2}$ times a resummation in powers of $\ln{Q^2/Q_T^2}$, to all orders in
$\alpha_s$.  It can be expressed as
\begin{equation}
\tilde{W}_{gg}(b,Q,x_{1},x_{2},C_{1,2,3})\ =\ e^{-S(b,Q,C_1,C_2)}\sum_{a,b}\bigl(C_{ga}\otimes f_a\bigr)(x_1)\bigl(C_{gb}\otimes f_b\bigr)(x_2)\,,
\end{equation}
where the Sudakov exponent is given by
\begin{eqnarray}
\label{sudakov}
S(b,Q,C_1,C_2)& = &\int_{C_1^2/b^2}^{C_2^2Q^2} \frac{d\bar{\mu}}{\bar{\mu}^2}\left[A(\alpha_s(\bar{\mu}),C_1)\ln\left(\frac{C_2^2 Q^2}{\bar{\mu}^2}\right)+B(\alpha_s(\bar{\mu}),C_1,C_2)\right]\,.
\end{eqnarray}
The coefficients $A$ and $B$ and the
Wilson coefficient functions $C_{ga}$ can be expanded as a power series in $\alpha_s$:
\begin{eqnarray}
A(\alpha_s(\bar{\mu}),C_1)& = & \sum_{n=1}^\infty \left(\frac{\alpha_s(\bar{\mu})}{\pi}\right)^n A^{(n)}(C_1), \\
B(\alpha_s(\bar{\mu}),C_1,C_2)&=& \sum_{n=1}^\infty \left(\frac{\alpha_s(\bar{\mu})}{\pi}\right)^n B^{(n)}(C_1,C_2),
\label{abfunction}
\end{eqnarray}
and
\begin{eqnarray}
C_{ga}(x,b,\mu,C_1,C_2)& = & \sum_{n=0}^\infty \left(\frac{\alpha_s(\mu)}{\pi}\right)^n C_{ga}^{(n)}(z,b,\mu,\frac{C_1}{C_2})\,,
\label{cfunction}
\end{eqnarray}
with $\mu=C_3/b$.
These quantities can be extracted order-by-order from the fixed-order calculations.  In our numerical results, we have included $A^{(1, 2, 3)}$,  $B^{(1, 2)}$ and $C^{(0, 1)}$,  whose analytical expressions are given in Appendix~\ref{sec:abc_functions} for completeness.  We use the canonical choice for the renormalization constants, $ C_1=C_3=2e^{-\gamma_E}$, $C_2=C_4=1$, which simplifies
the above expressions.
The function $\tilde{W}_{gg}^{NP}$ describes the non-perturbative part of the soft-gluon resummation, in which we use the
BLNY parameterization~\cite{Landry:2002ix}, but with the nonperturbative coefficients $g_{1,2,3}$ scaled by the factor
$C_A/C_F=9/4$. This scaling factor is to recognize that the initial state partons are gluons in the $gg \rightarrow H$ process,
in contrast to quarks in the Drelll-Yan process.

Finally, in Eq.~(\ref{eq:resum_formalism}), the term containing $Y$ incorporates the remainder of the cross section, which is not
singular as $Q_T\rightarrow 0$.
It consists of the difference between the full cross section at finite $Q_T$ and the small-$Q_T$ limit of this cross section, each calculated to the same order in $\alpha_s$.  At small $Q_T$,
these cancel, so that the contribution
of the $Y$-term is small, and the resummed term dominates.  At large $Q_T$, where the logarithms become small, the resummed term cancels against the small-$Q_T$ limit term (to the given order in $\alpha_s$), so that the cross section
approaches the fixed-order calculation.   More details of how this matching process between the resummed calculation and the fixed-order calculation is implemented in ResBos can be found in
Ref.~\cite{Balazs:1997xd}.  In previous studies of Higgs boson production at hadron colliders~\cite{Balazs:2000wv,Cao:2007du},
the high $Q_{T}$ perturbative calculation was included in ResBos at ${\cal O}(\alpha_{s}^{3})$.  The major update to the program that we have incorporated in this paper is to include the high $Q_{T}$ perturbative calculation at ${\cal O}(\alpha_{s}^{4})$.
This was done by using the code of Ref.~\cite{Glosser:2002gm} to rescale the perturbative piece of the grids used by the ResBos code by the factor (Pert($\alpha_{s}^3$)+Pert($\alpha_{s}^4$))/Pert($\alpha_{s}^3$), where Pert($\alpha_{s}^3$) and  Pert($\alpha_{s}^4$) are the contributions to the $Q_T$ distribution of the Higgs boson of order  $\alpha_{s}^3$ and $\alpha_{s}^4$, respectively.
Accordingly, the singular terms, called the aymptotical piece, at $O(\alpha_{s}^4)$, should also be included to ensure the cancellation
to Pert($\alpha_{s}^4$) piece in the low $Q_T$ region.

Note that the description of the perturbative piece in terms of LO,
NLO, or NNLO is problematic; for example, ${\cal O}(\alpha_{s}^{4})$
would be considered as NLO when referring to the (non-zero) transverse
momentum distribution, but it would be considered as NNLO when referring
to the total Higgs production cross section. Thus, we will refer to
the power of $\alpha_{s}$ when comparing the precision of the perturbative
piece of the calculation used in this work (${\cal O}(\alpha_{s}^{4})$)
versus that used in the previous works (${\cal O}(\alpha_{s}^{3})$).
Furthermore, both the  Pert($\alpha_{s}^3$)  and  Pert($\alpha_{s}^4$) contribnutions
are evaluated at the scale $C_4 Q$ with $C_4=C_2=1$ being the canonical choice of the
constants of the renormalization group equation for yielding the renormalization
formalism.

\subsection{Predictions of the total cross section for $gg\to HX$ at hardon colliders}
\label{sec:h_prod}
\begin{table}
\begin{tabular}{c|c|c|c|c|c|c}
\hline
 & \multicolumn{3}{c|}{Tevatron 1.96 TeV} & \multicolumn{3}{c}{LHC 7 TeV}\tabularnewline
\hline
$m_H$ (GeV) &\,\, RES \,\,& NLO$^{RES}$ & NLO$^{Higlu}$  & \,\, RES \,\, & NLO$^{RES}$ & NLO$^{Higlu}$ \tabularnewline
\hline
140 & 0.50 & 0.37 & 0.39 & 9.57 & 7.36   & 7.77 \tabularnewline
\hline
150 & 0.40 & 0.30 & 0.32 & 8.31 & 6.42   & 6.72 \tabularnewline
\hline
160 & 0.33 & 0.24 & 0.26 & 7.27 & 5.66   & 5.86 \tabularnewline
\hline
170 & 0.27 & 0.20 & 0.21 & 6.41 & 4.95   & 5.15 \tabularnewline
\hline
%
\hline
 & \multicolumn{3}{c|}{LHC 10 TeV} & \multicolumn{3}{c}{LHC 14 TeV}\tabularnewline
\hline
$m_H$ (GeV) &\,\, RES \,\,& NLO$^{RES}$ & NLO$^{Higlu}$  & \,\, RES \,\, & NLO$^{RES}$ & NLO$^{Higlu}$ \tabularnewline
\hline
140 & 18.1 & 14.4 & 15.2 & 32.1 & 25.8 & 27.2 \tabularnewline
\hline
150 & 15.9 & 12.7 & 13.3 & 28.9 & 23.0 & 24.1 \tabularnewline
\hline
160 & 14.1 & 11.3 & 11.7 & 25.6 & 20.6 & 21.5 \tabularnewline
\hline
170 & 12.5 & 10.1 & 10.5 & 23.0 & 18.6 & 19.3 \tabularnewline
\hline
\end{tabular}
\caption{Production cross section of the SM Higgs boson via the gluon fusion process, $gg \to HX$,  in \emph{pb} at the LHC and Tevatron. We  show the results for 7 TeV, 10 TeV, and 14 TeV c.m.~energy at the LHC and 1.96 TeV c.m.~energy at the Tevatron.
The label RES indicates the results of resummation calculations predicted from ResBos, and the labels NLO$^{RES}$ and NLO$^{Higlu}$ indicate the NLO results calculated from ResBos and Higlu codes, respectively.~\label{tab:xsection}}
\end{table}

The primary use of the resummation code ResBos is for the calculation of the transverse momentum spectrum, as
well as other distributions that are influenced strongly by soft gluon effects.  However, it also gives a calculation of the total cross section that is comparable to that of a fixed-order calculation, depending on the order to which the resummation
coefficients have been included.  We have included all of the coefficients ($A^{(1)}$, $B^{(1)}$, $C^{(0,1)}$) in ResBos
that are necessary to produce a NLO calculation of the cross section.  In addition, we also have included the NNLO
coefficients $A^{(2)}$ and $B^{(2)}$ and are only missing the function $C^{(2)}$ that is necessary to give a NNLO
calculation of the total
cross section.  The function $C^{(2)}$ should be extractable from the NNLO analytic expression of the cross section,
but this has not been achieved as yet.
To improve the resummation calcaultion, we also include  $A^{(3)}$ in the Sudakov exponent.
Thus, our calculation of the cross section should be comparable to a NLO
fixed-order calculation, and in fact contains much of the (logrithmic) contributions at NNLO.

In the remainder of this section we compare the predictions for the total cross section from ResBos against
NLO predictions.  In Table~\ref{tab:xsection}, we present the total cross sections of $gg\to HX$
for several benchmark points from the resummation (RES) calculations
using the updated ResBos program with CTEQ6.6M Parton Distribution Functions (PDFs)~\cite{Nadolsky:2008zw}.  These
are compared to an expansion of the resummation formula to NLO (NLO$^{RES}$), and
also to an exact NLO calculation.
The former is an exact NLO QCD calculation with the same implementation for including the effect
of the masses of the top, bottom, and charm quarks at LO, cf. Eq.~(3).
The latter is calculated with the help of the public code HIGLU~\cite{Spira:1995mt}.%
\footnote{More detailed analysis of the HIGLU calculation, e.g. the PDF uncertainties and
scale dependence, are given in the Appendix~\ref{sec:NLO}.
Unless specified otherwose, we only include teh exact NLO QCD contribution from HIGLU calulations.
}
We note that the main difference between these two NLO calculations is in the handling of the
quark mass dependences.
As explained in section~\ref{sec:css}, the resummed calculation, as well as the NLO$^{RES}$ calculation,
 is performed in the heavy top quark mass limit,
with the quark mass dependence included by the LO factor $\kappa$ given in Eq.~(\ref{eq:kappa}).
On the contrary, HIGLU uses the exact NLO two-loop calculation, including both the top quark and the bottom quark
in the loop.

\begin{figure}
\includegraphics[scale=0.62,]{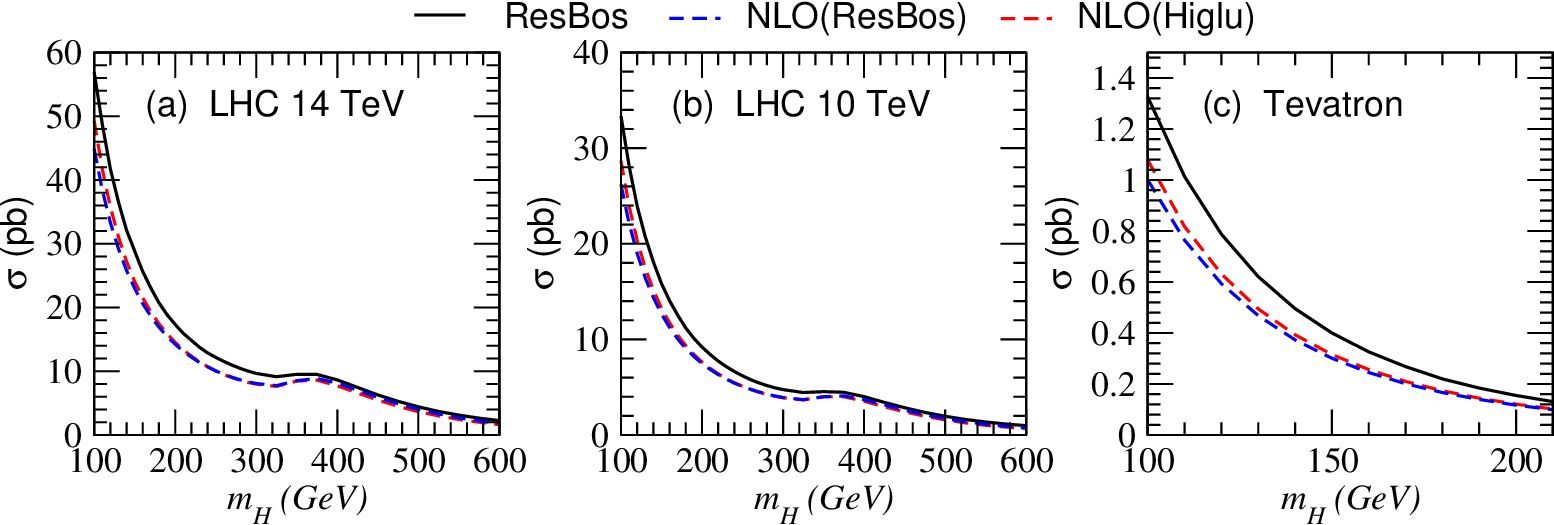}
\caption{The total cross section for $gg\to HX$ at the LHC and the Tevatron
using the updated ResBos program with CTEQ6.6M PDFs.
We consider 7 TeV, $10$ TeV and $14$ TeV c.m.~energy at the LHC and $1.96$ TeV c.m.~energy at Tevatron.
The NLO predictions from ResBos and Higlu programs are also shown.}
\label{fig:RESvsNLO}
\end{figure}

\begin{figure}
\includegraphics[scale=0.6,]{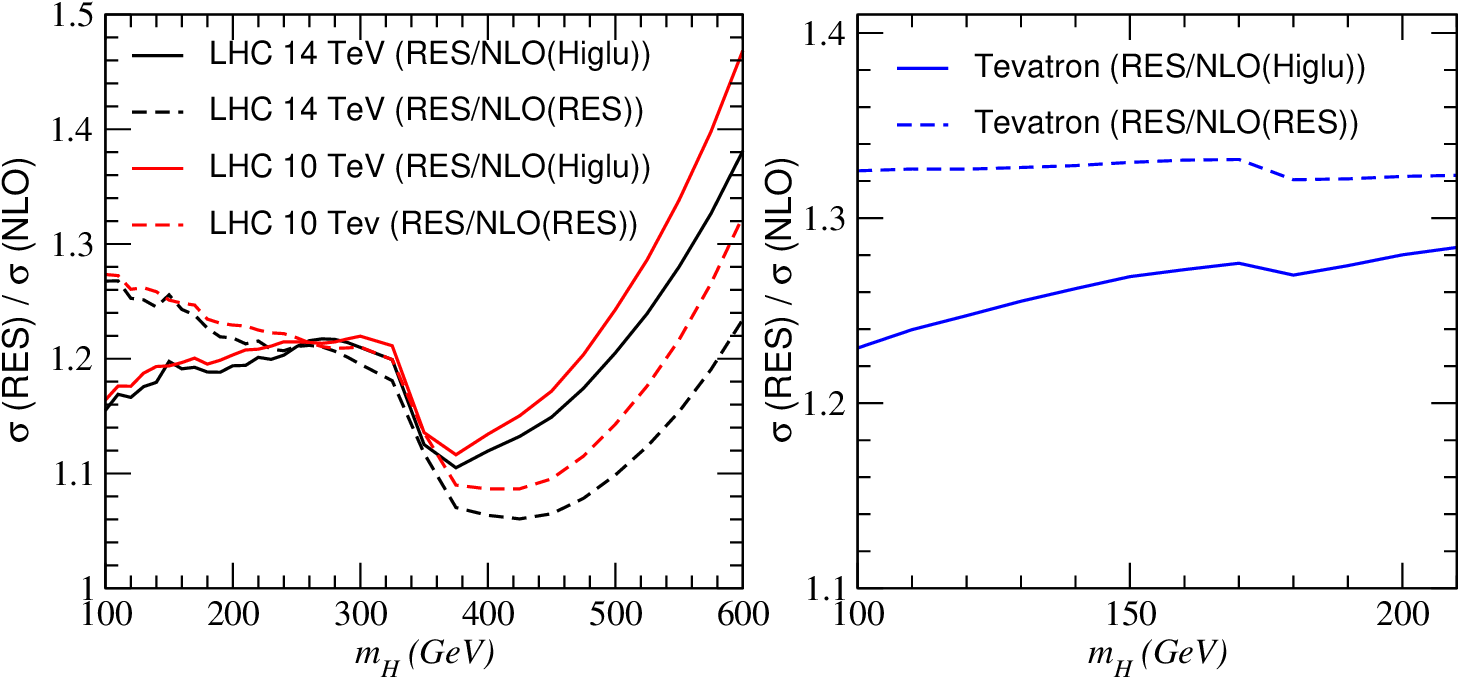}
\caption{The ratio of the total cross section of $gg\to H$ predicted from the resummation calculation
to that from NLO calculations at the 7 TeV, 10 TeV and 14 TeV LHC and the Tevatron.~\label{fig:ratio}}
\end{figure}

The cross section is consistently higher in the resummed calculation (RES) than for the
NLO calculations, due to the enhancement from the NNLO corrections.  This is seen more
easily in Fig.~\ref{fig:RESvsNLO}, which displays the cross section of the Higgs boson production
as a function of $m_H$ at the LHC and Tevatron, where the  (black) solid,  (blue) dashed and (red)  dotted curves
denote the RES, NLO$^{RES}$ and HIGLU results, respectively.  We display explicitly the enhancement
in the cross section in the resummed calculation in Fig.~\ref{fig:ratio},
where we plot the ratio of the RES cross section to the NLO cross section, both for the NLO$^{RES}$ and the
HIGLU calculations.
The ratio of RES to NLO drops rapidly when $m_H \gtrsim 300~\rm{GeV}$ and reaches a minimum
for  $m_H \sim 380~\rm{GeV}$.
This unusual dependence on $m_H$ can be traced to the handling of the virtuality of the Higgs boson in
the calculations.  The HIGLU calculation is for an explicitly on-shell Higgs boson production
without the subsequent Higgs boson decay, i.e. $\delta(Q^2-m_H^2)$. The same is also true for the
NLO$^{RES}$ calculation. On the contrary, the ResBos code takes into account the
Breit-Wagner width effects, see Eq.~(\ref{eq:resum_formalism}), to perform a realistic simulation.  Thus, the
cross section is calculated for an off-shell Higgs boson of mass-squared $Q^2$, which is then convoluted
with the Breit-Wigner.  This gives a sizable effect when the Higgs width is large, and in particular for
masses around 300-400 GeV  ($m_H \gtrsim 2 m_t$) where the cross section curves have noticeable structure.

\section{Numerical study of $gg \to H \to W W^{(*)} \to \ell\ell\nu\nu$}
\label{sec:numerical}

For our analysis of kinematic distributions, we will use several benchmark values for the Higgs boson mass:  $m_{H}=140,\,150,\,160$ and $170$
GeV at the Tevatron RUN II, and $m_{H} = 160$ GeV at the LHC.   We will focus on the most promising discovery mode for
this mass range, which is $gg\to H\to W^{+}W^{-}\to\ell^{+}\ell^{-}\nu\bar{\nu}$,
where $\ell^{\pm}$ denotes a charged lepton and $\nu(\bar{\nu})$ is a neutrino (anti-neutrino).
Note that all of the cross sections given in this section include the decay branching ratio of the Higgs boson and that the flavors of leptons are not summed over.
(Namely, only one lepton flavor, say, electron, is included here.)
\begin{figure}
\includegraphics[scale=0.6]{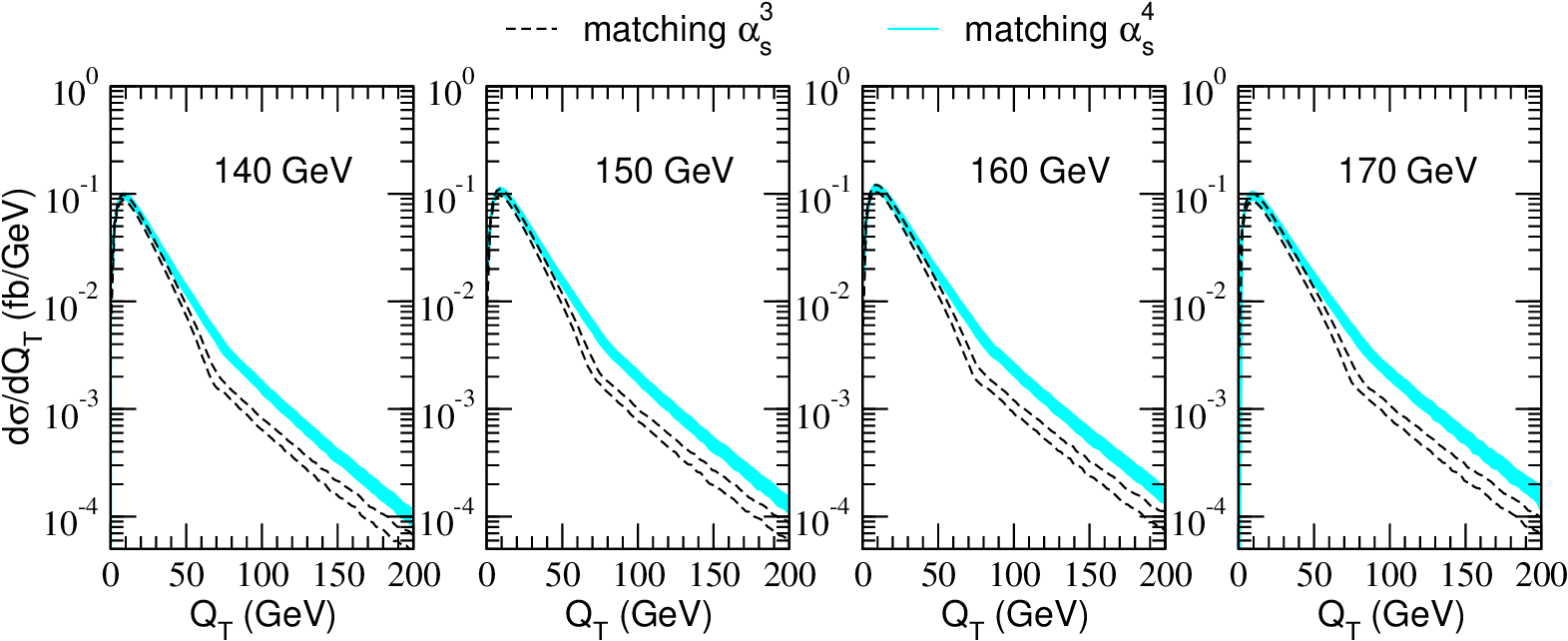}

\caption{
Transverse momentum, $Q_T$, of the SM Higgs boson at the Tevatron.
The black dashed lines and cyan lines are the calculations when matching at high $Q_T$ to the ${\cal O}(\alpha_{s}^{3})$ and ${\cal O}(\alpha_{s}^{4})$ fixed-order perturbative contributions, respectively, and the bands reflect the PDF uncertainties..~\label{fig:pt_tev}}
\end{figure}

\begin{figure}
\includegraphics[scale=0.55]{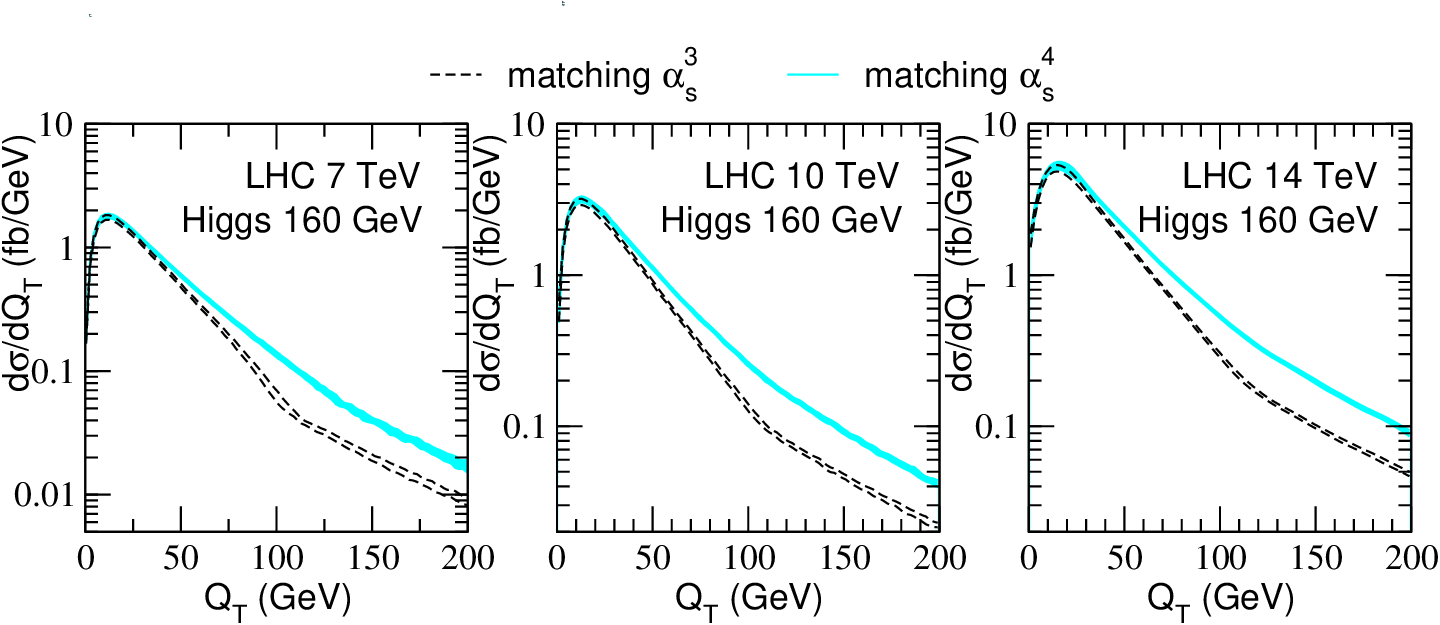}

\caption{
The same as Fig.~\ref{fig:pt_tev}, but for the SM Higgs boson at the 7 TeV, 10 TeV and 14 TeV LHC.}
\label{fig:pt_lhc}
\end{figure}

We begin with the transverse momentum distribution of the Higgs boson at the Tevatron, displayed in Fig.~\ref{fig:pt_tev}. The cyan lines and
the black dashed lines are the predictions with matching at
high $Q_T$ to ${\cal O}(\alpha_{s}^{4})$ and  ${\cal O}(\alpha_{s}^{3})$ fixed-order perturabative
calculations, respectively, and the bands show the uncertainties induced by
the eigenvector sets of CTEQ6.6 PDFs~\cite{Nadolsky:2008zw}.
The peak position (at $Q_{T}\sim10\,\rm{GeV}$) is
the same for both the dashed and cyan lines, since this is determined exclusively
by the resummed contribution to the calculation. At high $Q_{T}$,
however, the distribution is determined mainly by the perturbative
contribution, and it is substantially higher for the ${\cal O}(\alpha_{s}^{4})$
calculation. We also note that the ${\cal O}(\alpha_{s}^{4})$ perturbative
contribution matches much more smoothly with the resummed contribution.
This is particularly apparent when plotted with a logarithmic scale, as in  Fig.~\ref{fig:pt_tev},
where the unphysical kink around $Q_{T}\sim70$ GeV in the ${\cal O}(\alpha_{s}^{3})$
curve is absent in the ${\cal O}(\alpha_{s}^{4})$ curve.
Similarly, we show the improved $Q_T$ distribution of the Higgs boson at the LHC in Fig.~\ref{fig:pt_lhc}, including PDF uncertainties as well. The kink seen in the ${\cal O}(\alpha_{s}^{3})$ dashed line is shifted to higher $Q_T$ ($\sim110$ GeV) compared with the case at the Tevatron. With matching to the ${\cal O}(\alpha_{s}^{4})$ calculation, the curve becomes much more smooth.
The uncertainty induced by the PDFs at the peak region is about $8\%$ and increases to higher than $10\%$ when $Q_T$ is larger than $40$ GeV at the Tevatron, while the uncertainty stays at about $2\%\sim 6\%$ for
the full range of $Q_T$ at the LHC.

\begin{figure}
\includegraphics[scale=0.55]{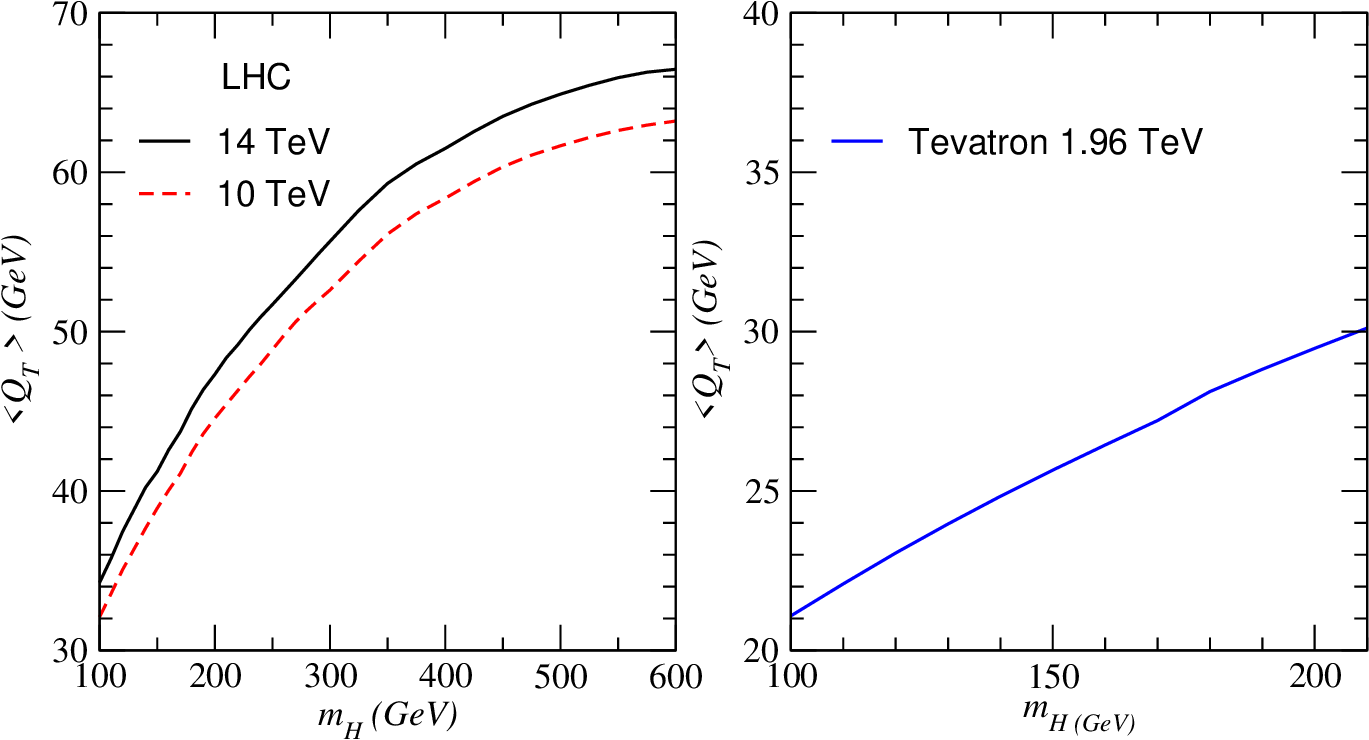}

\caption{The average of $Q_T$ (in the region of $0\le Q_T\le 200$ GeV) of the Higgs boson at the
7 TeV, 10 TeV and 14 TeV LHC and the Tevatron.~\label{fig:pt_avg}}

\end{figure}

We can use the updated $Q_T$ distributions at the LHC and the Tevatron to study the dependence of this distribution on the collider
c.m.~energy.   The peak position, $Q_T\rm{(peak)}$, only changes mildly with the c.m.~energy.
For example, $Q_T\rm{(peak)}$ is roughly $10$ GeV,  $10$ GeV, $13$ GeV and $13$ GeV at the Tevatron, LHC 7 TeV,
LHC $10$ TeV and LHC $14$ TeV c.m.~energy,
respectively, for $m_H = 160$ GeV. On the other hand, the average transverse momentum, $\langle Q_T\rangle$, increases more substantially
when the mass of the Higgs boson or the energy of the collider increases.
For $m_H=160$ GeV and taking the average in the region of $0\le Q_T \le 200$ GeV, we find that $\langle Q_T\rangle$ increases from about $26$ GeV
at the Tevatron to about $40$ GeV, $40$ GeV and $43$ GeV at the LHC  with 7 TeV, $10$ TeV and $14$ TeV c.m.~energy, respectively. The dependence of $\langle Q_T\rangle$ on the Higgs boson mass at the various colliders is shown in Fig.~\ref{fig:pt_avg}.

\begin{figure}[t]
\includegraphics[scale=0.55,]{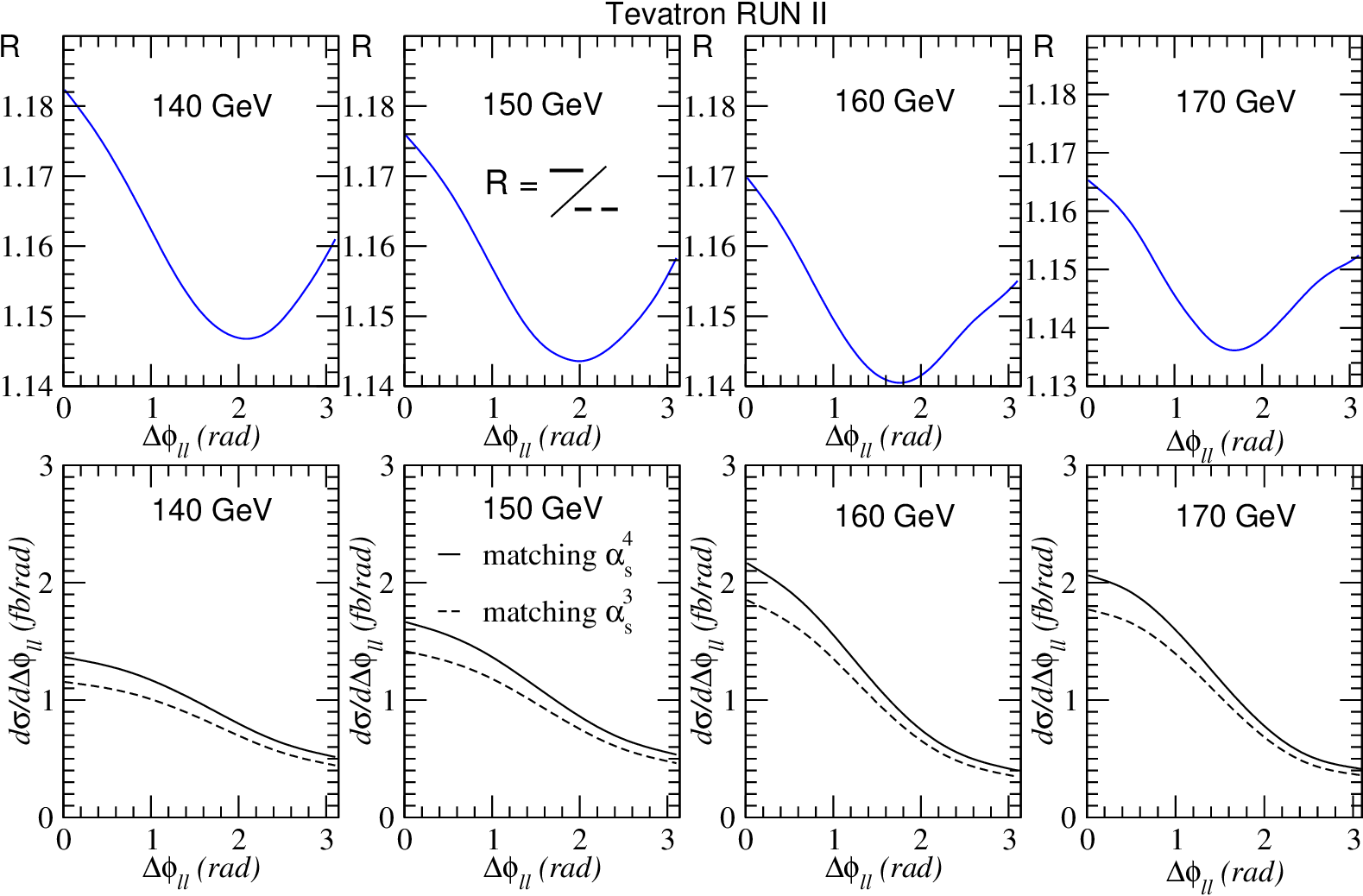}

\caption{$\Delta\phi_{\ell\ell}$ distributions at the Tevatron.
The solid and the dashed curves in the lower panel are the calculations when
matching at high $Q_T$ to the ${\cal O}(\alpha_{s}^{4})$ and ${\cal O}(\alpha_{s}^{3})$
fixed-order calculations , respectively.
The blue curve in the upper panel is the ratio of the solid curve to the
dashed curve.~\label{fig:dph_tev}}

\end{figure}

\begin{figure}
\includegraphics[scale=0.5,]{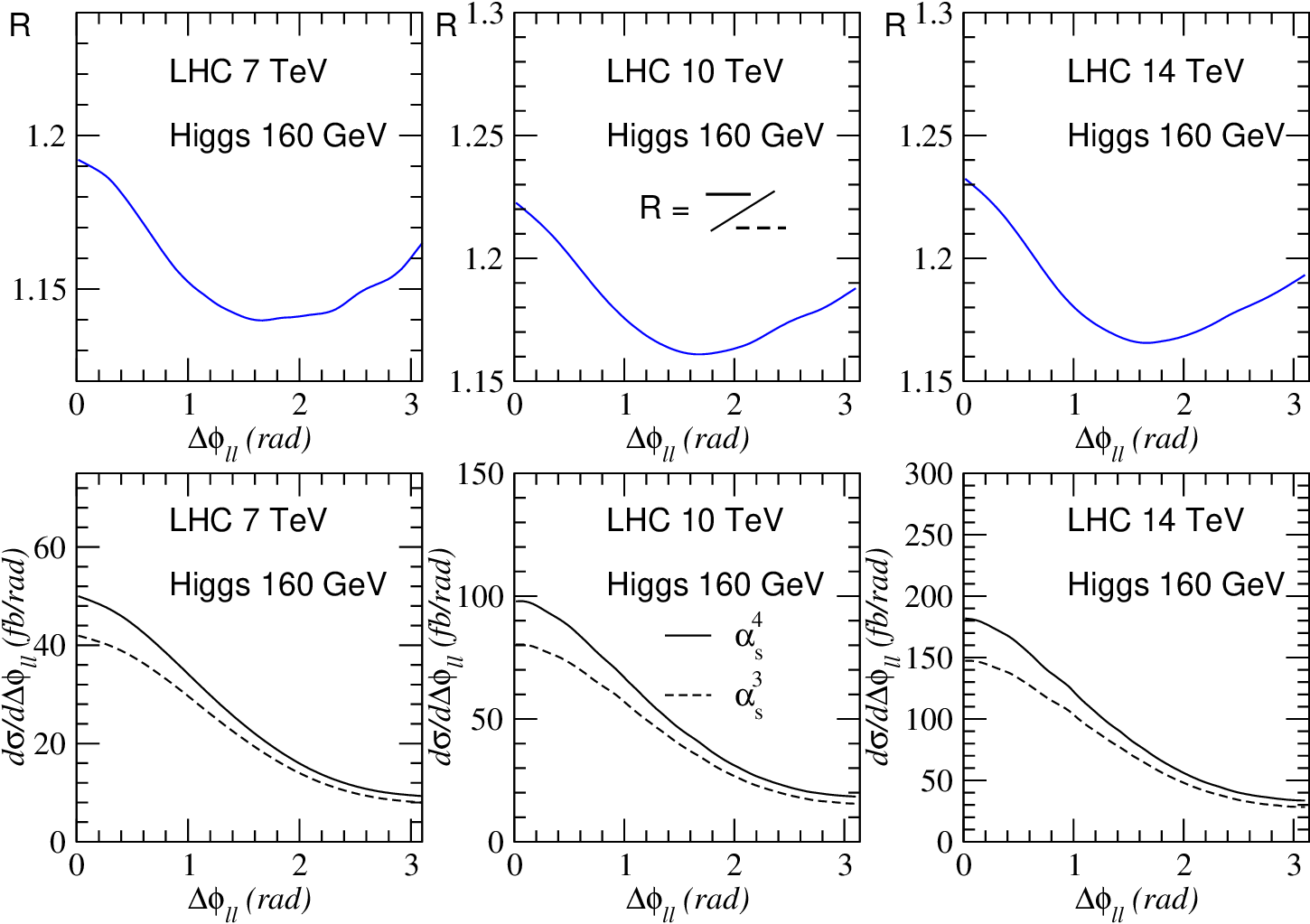}

\caption{The same as Fig.~\ref{fig:dph_tev} but for the 7 TeV, 10 TeV and 14 TeV LHC.~\label{fig:dph_lhc}}

\end{figure}

Finally, we consider the distribution in the difference in azimuthal angle of the two charged leptons in the Higgs boson decay, $\Delta\phi_{\ell\ell}$, which is  useful for making cuts to extract the Higgs boson signal from background.  This distribution is shown for the Tevatron in Fig.~\ref{fig:dph_tev}.  The solid and dashed lines in the lower panel are the calculations when matching to  ${\cal O}(\alpha_{s}^{4})$ and ${\cal O}(\alpha_{s}^{3})$ at high $Q_T$, respectively.  From these plots we see that the signal process
$gg\to H\to W^{+}W^{-}\to\ell^{+}\ell^{-}\nu\bar{\nu}$ prefers to have the charged leptons both moving in the same direction,
i.e., $\Delta\phi_{\ell\ell}\approx0$.   This can be understood from angular momentum conservation and the left-handed nature
of the $W$-boson decays.  In the Higgs boson rest frame, the $W^+$ and $W^-$ are produced back-to-back with opposite polarizations.  The $W^-$ boson decays with the charged lepton momentum
anti-correlated with the $W^-$ polarization, while the $W^+$ boson decays with the charged
anti-lepton momentum correlated with the
$W^+$ polarizarion.  As a result the two charged leptons tend to move in the same direction. This feature still holds even when only one of the $W$ bosons from the Higgs boson decay is on-shell. On the other
hand, the background events, which predominantly originate from $W^+W^-$ pair production, are more likely to be produced with the charged leptons back-to-back with $\Delta\phi_{\ell\ell}\approx\pi$.

In the second row of  Fig.~\ref{fig:dph_tev}, we show the $d \sigma/d \Delta \phi_{ll}$
cross sections when matching at high $Q_T$ to the
${\cal O}(\alpha_{s}^{4})$ (solid curve) and ${\cal O}(\alpha_{s}^{3})$ (dashed curve),
fixed-order calculations, respectively. The upper panel shows the ratio of
these two predictions.
The enhancement of the distribution due to the improved calculation is evident.
We see that the improved ResBos calculation with matching at ${\cal O}(\alpha_{s}^{4})$ enhances $\Delta\phi_{\ell\ell}$  by about from $14\%$ to $18\%$ over the old calculation.  Interestingly, the largest enhancement occurs at $\Delta\phi_{\ell\ell}\approx0$, while the
enhancement has a minimum around $\Delta\phi_{\ell\ell}\approx2$ radians.
Similarly, in the $\Delta\phi_{\ell\ell}$ distribution at the LHC, which is shown in Fig.~\ref{fig:dph_lhc}, the enhancement due to the improved matching varies between $16\% \sim 23\%$, which is slightly larger than for the Tevatron.
The change in the shape of the $\Delta\phi_{\ell\ell}$ distribution between the ${\cal O}(\alpha_{s}^{3})$ and the ${\cal O}(\alpha_{s}^{4})$ calculations can be considered purely kinematical,
since the Higgs boson is a scalar and the corresponding distribution in the Higgs boson rest
frame would not be affected by changes in the Higgs boson
production cross section.  In the lab frame, however, the distribution is affected by the fact that the
cross section is larger at high Higgs boson $Q_T$ in the ${\cal O}(\alpha_{s}^{4})$ calculation, so that more of the charged lepton pairs are produced with a bigger momentum boost.
\section{Scale Dependence}
\label{sec:scale_dep}

Up to now we have taken the canonical choice of the renormalization and factorization scales,
specified by the renormalization constants $C_{i}$, for $i=1,2,3,4$, in the resummation calculation.
It is desirable to vary some of the scales in the resummation formalism to examine the effects of
scale dependences on various kinematical distributions in the Higgs boson production and decay. 
The $Q_T$ distribution of the Higgs boson is particularly important as it is used to 
model the Higgs boson $p_T$ spectrum at the Tevatron~\cite{Peters:2010tv}. 
The usual practice to estimate the size of the yet-to-be calculated higher order contributions 
is to vary the hard scales by a factor of two around the typical hard scale of the considered process. 
We vary the renormalization constants $C_{i}$ around their canonical values by a factor of two, but with the relations
that $C_{1}=C_{2}b_{0}$, $C_{3}=b_{0}$ and $C_{4}=C_{2}$, for a varying $C_{2}=2$, $1$ and $0.5$.
With the choice of $C_{i}$, the Wilson coefficient functions $C_{gg}$ and $C_{gq}$
are not altered; see Eq.~\ref{eq:C_function}. 
Hence, the dominant effect of the variation is to change the shape, but not the rate, of
various kinematical distributions of Higgs boson produced via gluon fusion process.

The total cross sections of $gg\rightarrow HX$ predicted from the
resummation calculations vary about $10\%$ for different choices of $C_{i}$ at various colliders; 
see Table~\ref{tab:rate-ci}. Decreasing $C_{2}$ enhances the total cross sections.

\begin{table}
\begin{tabular}{|c|c|c|c|c|c|c|}
\hline
 & \multicolumn{3}{c|}{Tevatron 1.96 TeV} & \multicolumn{3}{c|}{LHC 7 TeV}\tabularnewline
\hline
$m_{H}$ (GeV) & $C_{2}=2$ & $C_{2}=1$ & $C_{2}=0.5$ & $C_{2}=2$ & $C_{2}=1$ & $C_{2}=0.5$\tabularnewline
\hline
140 & 0.47 & 0.51 & 0.59 & 9.07 & 9.57 & 10.4 \tabularnewline
\hline
150 & 0.38 & 0.41 & 0.47 & 7.89 & 8.31 & 9.05\tabularnewline
\hline
160 & 0.31 & 0.33 & 0.38 & 6.92 & 7.27 & 7.93\tabularnewline
\hline
170 & 0.26 & 0.27 & 0.32 & 6.11 & 6.41 & 6.99\tabularnewline
\hline
%
\hline
 & \multicolumn{3}{c|}{LHC 10 TeV} & \multicolumn{3}{c|}{LHC 14 TeV}\tabularnewline
\hline
$m_{H}$ (GeV) & $C_{2}=2$ & $C_{2}=1$ & $C_{2}=0.5$ & $C_{2}=2$ & $C_{2}=1$ & $C_{2}=0.5$\tabularnewline
\hline
140 & 17.6 & 18.5 & 20.2 & 31.4 & 33.0 & 35.9 \tabularnewline
\hline
150 & 15.5 & 16.3 & 17.7 & 28.0 & 29.2 & 31.8 \tabularnewline
\hline
160 & 13.8 & 14.4 & 15.7 & 25.1 & 26.1 & 28.4 \tabularnewline
\hline
170 & 12.3 & 12.8 & 14.0 & 22.6 & 23.5 & 25.5 \tabularnewline
\hline
\end{tabular}
\caption{\label{tab:rate-ci}Total cross sections (pb) of the SM Higgs production via gluon fusion for different scale choices
at the Tevatron and the LHC. These results are predicted from the resummation calculations using ResBos.}
\end{table}

\begin{figure}
\includegraphics[width=0.3\textwidth]{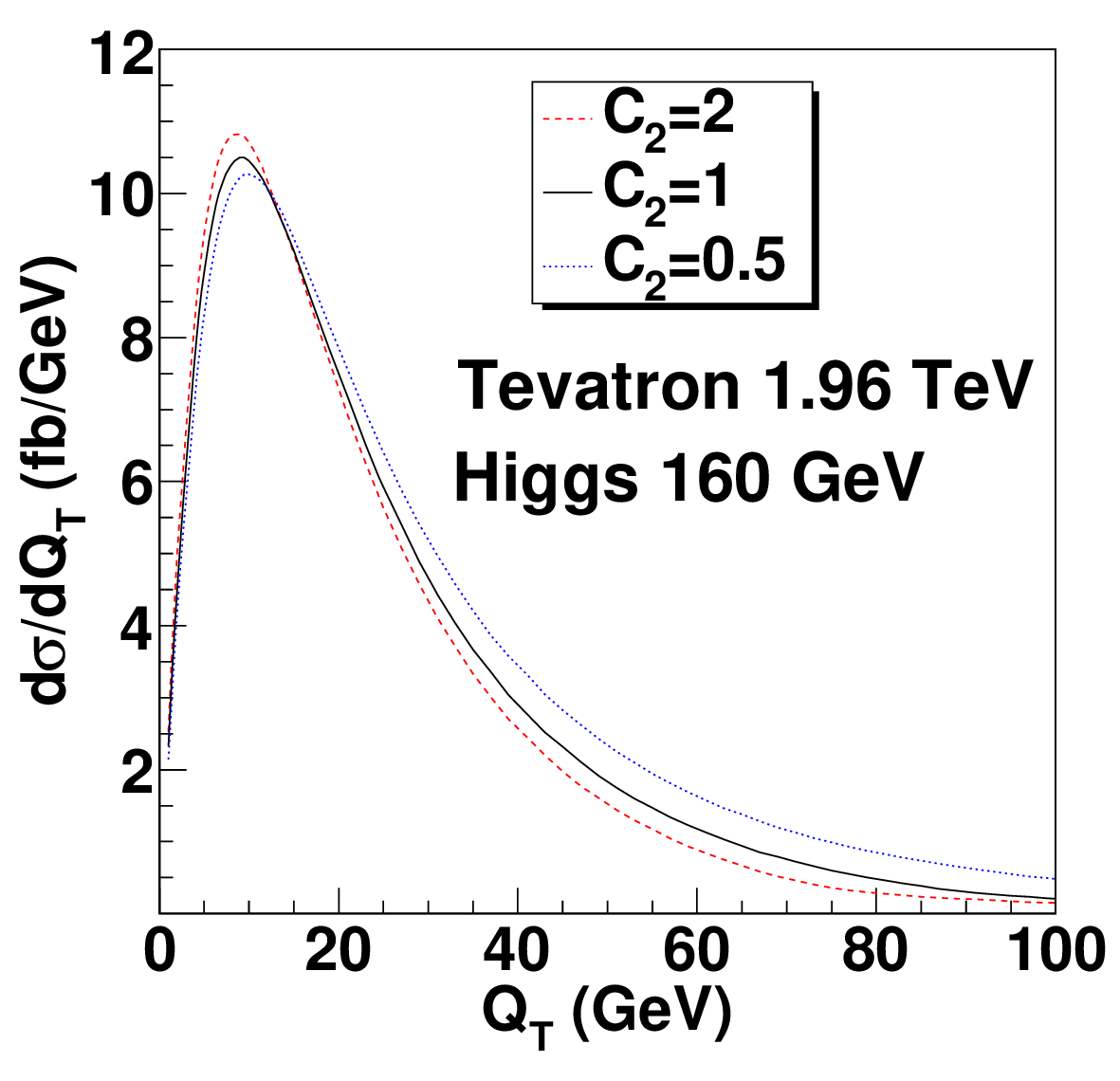}
\includegraphics[width=0.3\textwidth]{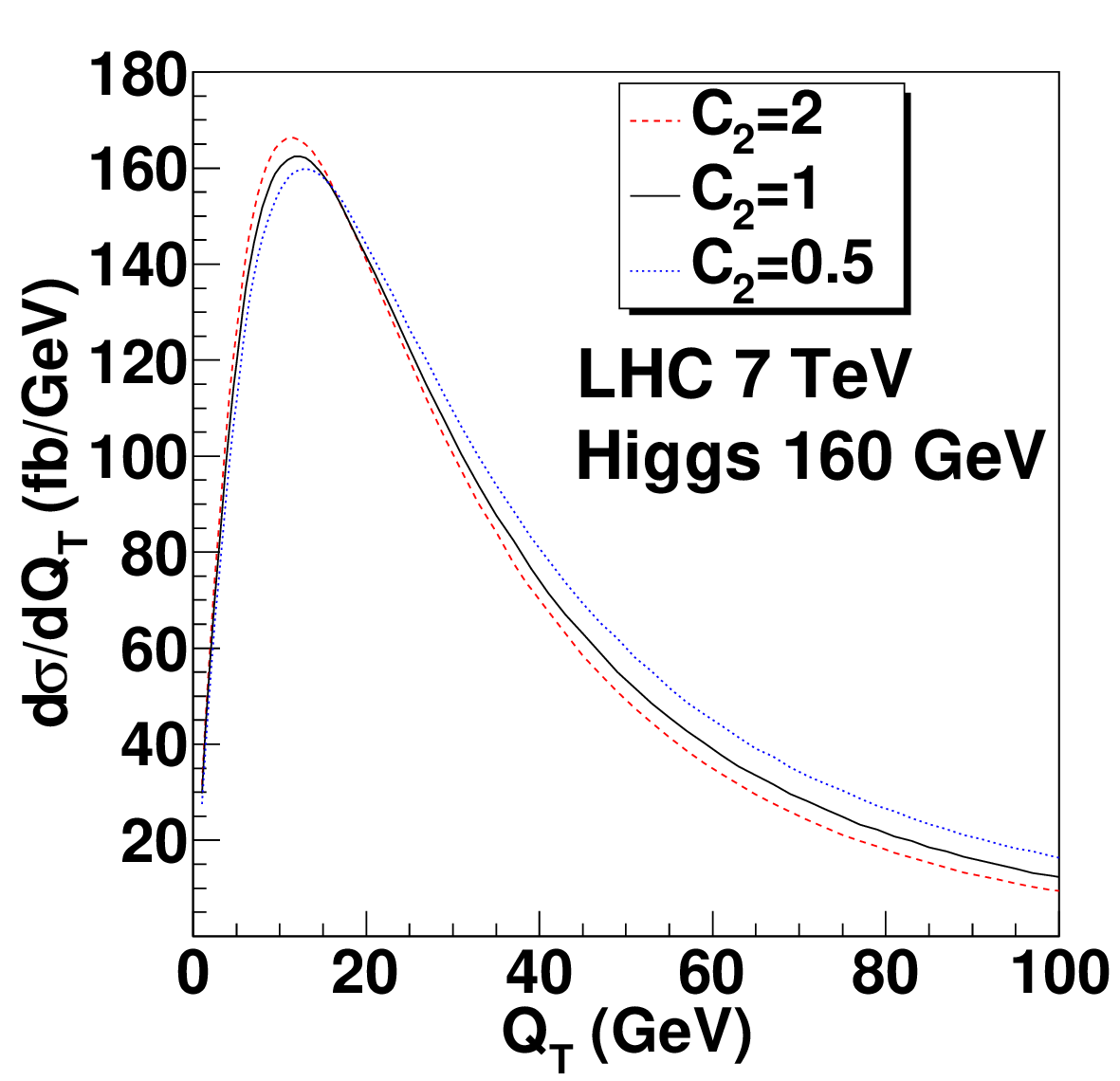}\\
\includegraphics[width=0.3\textwidth]{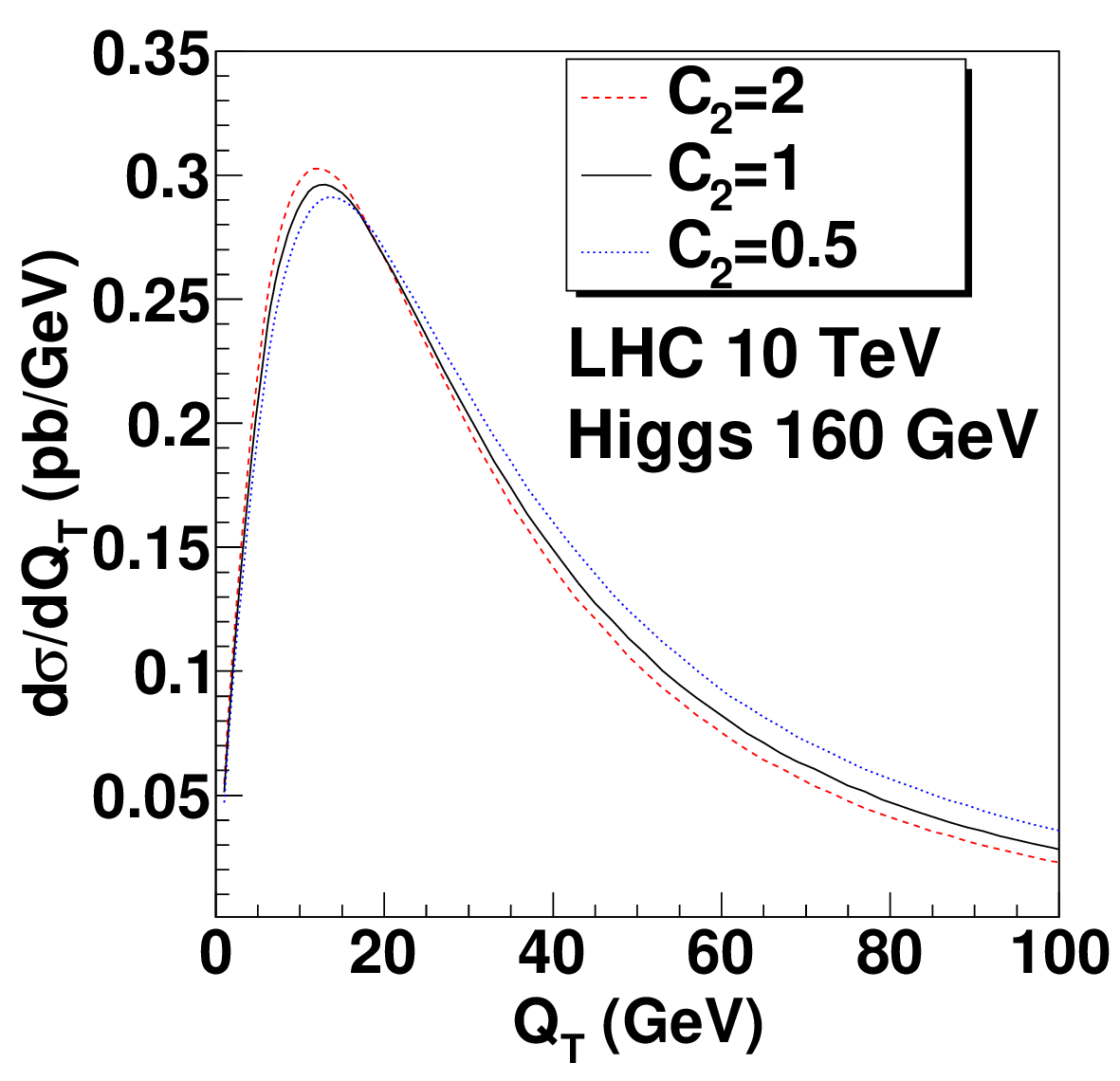}
\includegraphics[width=0.3\textwidth]{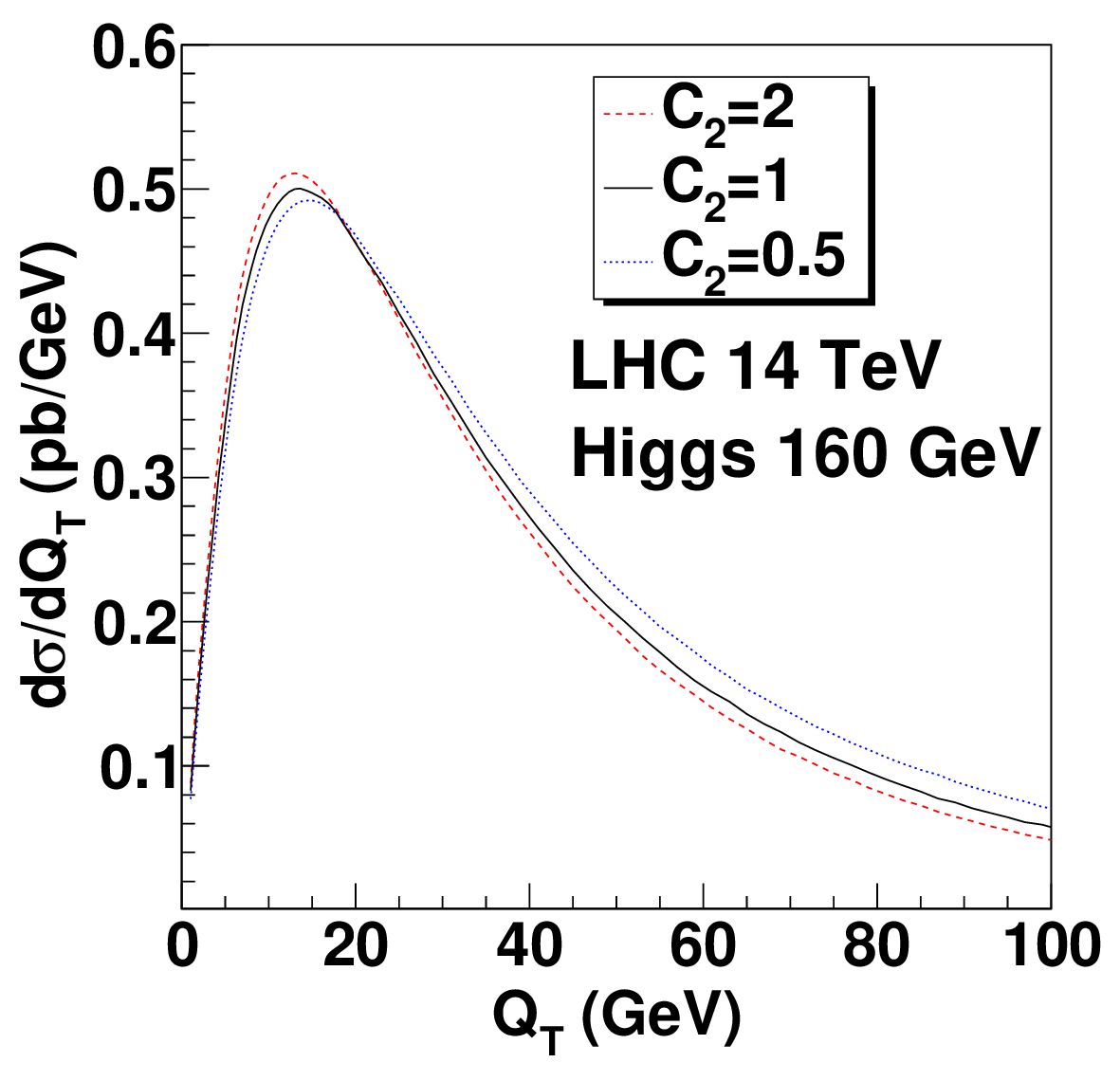}
\caption{
Transverse momentum $Q_T$ distribution of the SM Higgs boson produced via gluon fusion
for variouis scale choices at the Tevatron and the LHC.
}
\label{fig:c2}
\end{figure}

The transverse momentum distributions of Higgs boson predicted
from resummation calculations using RESBOS are shown in Fig.~\ref{fig:c2}.
The shapes of the transverse momentum distribution of Higgs boson at
various colliders are changed.
Especially, the peaks are shifted by a few GeV as varying
the scales. In general, increasing $C_{2}$ enhances the
low $Q_{T}$ region but suppress the high $Q_{T}$ region. The  $Q_{T}$ distributions
for different scale choices cross at $15$ GeV ($20$ GeV) at the
Tevatron (LHC). The main contribution in the low $Q_T$ region, which
dominates the total cross section, comes from the resummation piece, which is scale
invariant after including enough high order calculations in $A$, $B$ and $C$ functions.
The difference
in the total cross sections originates mainly from the fixed-order contributions,
which are matched with resummation piece to obtain the physical distributions.

\section{Conclusion}
\label{sec:conclusion}
The search for the Higgs boson at the Tevatron and the LHC relies on detailed knowledge of the distributions (event rate and shape) of the Higgs decay products.  These, in turn, are sensitive to the kinematics of the Higgs boson production. In this paper, we improve the calculation of the transverse momentum, $Q_T$, distribution of the Higgs boson in the gluon fusion production process, $gg\to H$, at the Tevatron and the LHC by matching the resummed distribution at small $Q_T$ with the ${\cal O}(\alpha_{s}^4)$ fixed-order perturbative calculation at high $Q_T$ in the ResBos Monte Carlo program.
The total cross section of $gg \to HX$ predicted from the updated ResBos is always larger than NLO calculation and the enhancement can reach about $40\%$ at the LHC for a heavy Higgs boson. The difference between RES and NLO calculation becomes  minimum ($\sim 10\% $) in the threshold regime of  $m_H\gtrsim 2m_t$.
The unphysical kink, which had been obtained when matching at high $Q_T$ with the ${\cal O}(\alpha_{s}^3)$ fixed-order perturbative calculation, is removed.
The PDF uncertainties are studied in the $Q_T$ distributions as well.  At the Tevatron we find the uncertainty to be about $5\%$ in the peak area and larger than $10\%$ in the region of $Q_T\gtrsim50$ GeV.
At the LHC, the uncertainty is about $2\%\sim6\%$ for the entire region of $Q_T$.
For the comparison of $Q_T$, we found that the average value of $Q_T$ increases when the c.m.~energy or the mass of the Higgs boson increases, however, peak position is insensitive to both.
We finally study the opening azimuthal angle , $\Delta\phi_{\ell\ell}$, of the two charged leptons in the
Higgs boson decay, $H\to W^{+}W^{-}\to\ell^{+}\ell^{-}\nu\bar{\nu}$.  The distribution of
$\Delta\phi_{\ell\ell}$ is enhanced by $14\%\sim18\%$ and $16\%\sim 23\%$ at the Tevatron and the LHC, respectively, with the largest enhancement occurring at small opening angles.  With current and expected integrated luminosity, the Tevatron has the capability to discover or exclude the SM Higgs boson over a significant range of mass. Detailed predictions of the kinematics of the Higgs boson, such as its transverse momentum distribution, play a crucial role in this
analysis.  This is our motivation for the continued improvement of the ResBos program and for this study.
\begin{acknowledgments}
Q.H.C. is supported in part by the Argonne National Laboratory and University of Chicago Joint Theory Institute (JTI) Grant 03921-07-137, and by the U.S. Department of Energy under Grants No. DE-AC02-06CH11357 and No. DE-FG02-90ER40560. C.R.C is supported by World Premier International Research Center Initiative (WPI Initiative), MEXT, Japan.  C.S. and C.P.Y. acknowledge the support of the U.S. National Science Foundation under grant PHY-0555544 and PHY-0555545.
C.P.Y. would also like to thank the hospitality of National Center for Theoretical Sciences in Taiwan and
Center for High Energy Physics, Peking University, in China, where part of this work was done.
C.P.Y acknowledges the support of the U.S. National Science Foundation under
Grand No. PHY-0855561.
\end{acknowledgments}

\appendix

\section{A, B, and C Coefficients}
\label{sec:abc_functions}
For completeness, we give expressions for the so-called $A$, $B$, and $C$ coefficients used in our resummation calculations. In our numerical result, we have used $A^{(1,2)}$, $B^{(1)}$~\cite{Catani:1988vd},
$B^{(2)}$~\cite{deFlorian:2000pr,deFlorian:2001zd}, $A^{(3)}$~\cite{Vogt:2004mw} and $C^{(0,1)}$~\cite{Kauffman:1991cx,Yuan:1991we}.
Their analytical expressions for the process $gg\to H$ are much simplified in the canonical choice of the renormalization constants,
$ C_1=C_3=2e^{-\gamma_E}$ and $C_2=C_4=1$,
which we use in this project, unless specified otherwise.  For this choice of renormalization constants, we have:
\begin{eqnarray}
 \label{eq:A_function}
 A_g^{(1)}& =  & C_A, \\
 A_g^{(2)}& = & C_A \left[\left(\frac{67}{36}-\frac{\pi^2}{12}\right)N_c-\frac{5}{18}N_f \right], \\
 A_g^{(3)}& = & \frac{C_A C_F N_f}{2}\left(\zeta(3)-\frac{55}{48}\right)-\frac{C_A N_f^2}{108}+C_A^3\left(\frac{11\zeta(3)}{24}+\frac{11\pi^4}{720}-\frac{67\pi^2}{216}+\frac{245}{96}\right)\nonumber\\
 & &+C_A^2 N_f \left(-\frac{7\zeta(3)}{12}+\frac{5\pi^2}{108}-\frac{209}{432}\right),
\end{eqnarray}
where $C_A = 3$, $N_c = 3$, $N_f = 5$, $C_F = 4/3$ and the Riemann constant $\zeta(3) = 1.202...$;
\begin{eqnarray}
 \label{eq:B_function}
 B_g^{(1)}& =  &- \beta_0, \\
 B_g^{(2)}& = &- \frac{1}{2}\left[C_A^2 \left(\frac{8}{3}+3\zeta(3)\right)-C_F T_R N_f -\frac{4}{3}C_A T_R N_f\right]+\beta_0\left[\frac{C_A \pi^2}{12}+\frac{11+3\pi^2}{4}\right] ,
\end{eqnarray}
where $\beta_0 = (11N_c-2N_f)/6$ and $T_R = 1/2$;
\begin{eqnarray}
 \label{eq:C_function}
 C_{gg}^{(0)}(x)& =  & \delta(1-x), \\
 C_{gq}^{(0)}(x)& =  &0, \\
 C_{gg}^{(1)}(x)& = &\delta(1-x)\frac{11+3\pi^2}{4} ,\\
 C_{gq}^{(1)}(x)& = & \frac{C_F}{2}x,
\end{eqnarray}
where $x$ is the momentum fraction carried by the gluon after splitting from its mother particle (gluon $g$ or quark $q$).

For non-canonical choice of the renormalization constants $C_{i}$, for $i=1,2,3,4$, extra terms are
needed to render the renormalization group invariance of the resummation formalism~\cite{Balazs:2007hr}.
To investigate the scale dependence of the kinematical distributions of Higgs boson
produced via gluon fusion process in this study,
we shall vary the constants $C_{i}$ with the following relations:
\begin{eqnarray}
 \label{eq:scale_choice}
C_{2} & = & 2,\ 1,\ 0.5 \, ,\\
C_{1} & = & C_{2}b_{0} ,\\
C_{3} & = & b_{0} ,\\
C_{4} & = & C_{2}.
\end{eqnarray}
With this choice, we have
\begin{eqnarray}
 \label{eq:general_scale}
\mathcal{A}_{g}^{(1)}(C_{1}) & = & \mathcal{A}_{g}^{(1,c)},\\
\mathcal{A}_{g}^{(2)}(C_{1}) & = & \mathcal{A}_{g}^{(2,c)}-\mathcal{\mathcal{A}}_{g}^{(1,c)}\beta_{0}\ln\frac{b_{0}}{C_{1}},\\
\mathcal{A}_{g}^{(3)}(C_{1}) & = & \mathcal{A}_{g}^{(3,c)}-2\mathcal{A}_{g}^{(2,c)}\beta_{0}\ln\frac{b_{0}}{C_{1}}-\frac{A_{g}^{(1,c)}}{2}\beta_{1}\ln\frac{b_{0}}{C_{1}}+\mathcal{A}_{g}^{(1,c)}\beta_{0}^{2}\left(\ln\frac{b_{0}}{C_{1}}\right)^{2},\\
\mathcal{B}_{g}^{(1)}(C_{1},C_{2}) & = & \mathcal{B}_{g}^{(1,c)},\\
\mathcal{B}_{g}^{(2)}(C_{1},C_{2}) & = & \mathcal{B}_{g}^{(2,c)}+\beta_{0}\left[\mathcal{A}_{g}^{(1,c)}\ln^{2}\frac{b_{0}}{C_{1}}+\mathcal{B}_{g}^{(1,c)}\ln C_{2}-\mathcal{A}_{g}^{(1,c)}\ln^{2}C_{2}\right],\\
\mathcal{C}_{gg}^{(1)}(x,b\mu,\frac{C_{1}}{C_{2}}) & = & \mathcal{C}_{gg}^{(1,c)}(x),\\
\mathcal{C}_{gq}^{(1)}(x,b\mu,\frac{C_{1}}{C_{2}}) & = & \mathcal{C}_{gq}^{(1,c)}(x),
\end{eqnarray}
where the superscript $c$ indicates the corresponding constant in the canonical case. For example,
$\mathcal{A}_{g}^{(1,c)}= A_g^{(1)}$, etc.
With this choice of $C_{i}$, the Wilson coefficient functions $C_{gg}$ and $C_{gq}$, cf. Eq.~\ref{eq:C_function},
are not altered. Hence, the dominant effect of the variation is to change the shape, not the rate, of
various kinematical distributions of Higgs boson produced via gluon fusion process.
 Moreover, when varying the hard scale $C_4 Q$, the Born level cross section $\sigma_0$, cf.
Eq.~\ref{eq:sigma0}, should be multiplied by the following factor
\begin{eqnarray}
 &  & [1+\frac{\beta_{0}}{2\pi}\alpha_{S}(Q)\ln C_{4}^{2}-\frac{\beta_{1}}{4\pi^{2}}\alpha_{S}^{2}(Q)\ln C_{4}^{2}]^{2}\\
 & = & 1+\frac{\beta_{0}}{\pi}\alpha_{S}(Q)\ln C_{4}^{2}+
 \frac{\alpha_{S}^{2}(Q)}{4\pi^{2}}[\beta_{0}^{2}\ln^{2}C_{4}^{2}-2\beta_{1}\ln C_{4}^{2}]+\mathcal{O}(\alpha_{S}^{3}).
\end{eqnarray}
They depend on the QCD beta-function coefficients $\beta_{0}=(11N_{c}-2N_{f})/6$,
$\beta_{1}=(17N_{c}^{2}-5N_{c}N_{f}-3C_{F}N_{f})/6$ for $N_{c}$
colors and $N_{f}$ active quark flavors, with $C_{F}=(N_{c}^{2}-1)/(2N_{c})=4/3$
for $N_{c}=3$.

\section{Total cross sections of $gg\to HX$ at NLO}
\label{sec:NLO}

In this section we evaluate the uncertainties in the cross section due to uncertainties in the PDFs
and due to higher order corrections, as illuminated by renormalization scale dependence.  Since we have
used the code HIGLU~\cite{Spira:1995mt} as a reference comparison for the total cross section in our
resummation calculation (see for example Fig.~\ref{fig:RESvsNLO}), we will use it to produce numerical
results in this section.  We expect the uncertainties
in the total cross section in the resummed calculation to be comparable.
Note that the Higgs boson decay is not implemented in HIGLU, so all the results presented in this appendix
are for on-shell Higgs boson production only.

\begin{figure}
\includegraphics[scale=0.4]{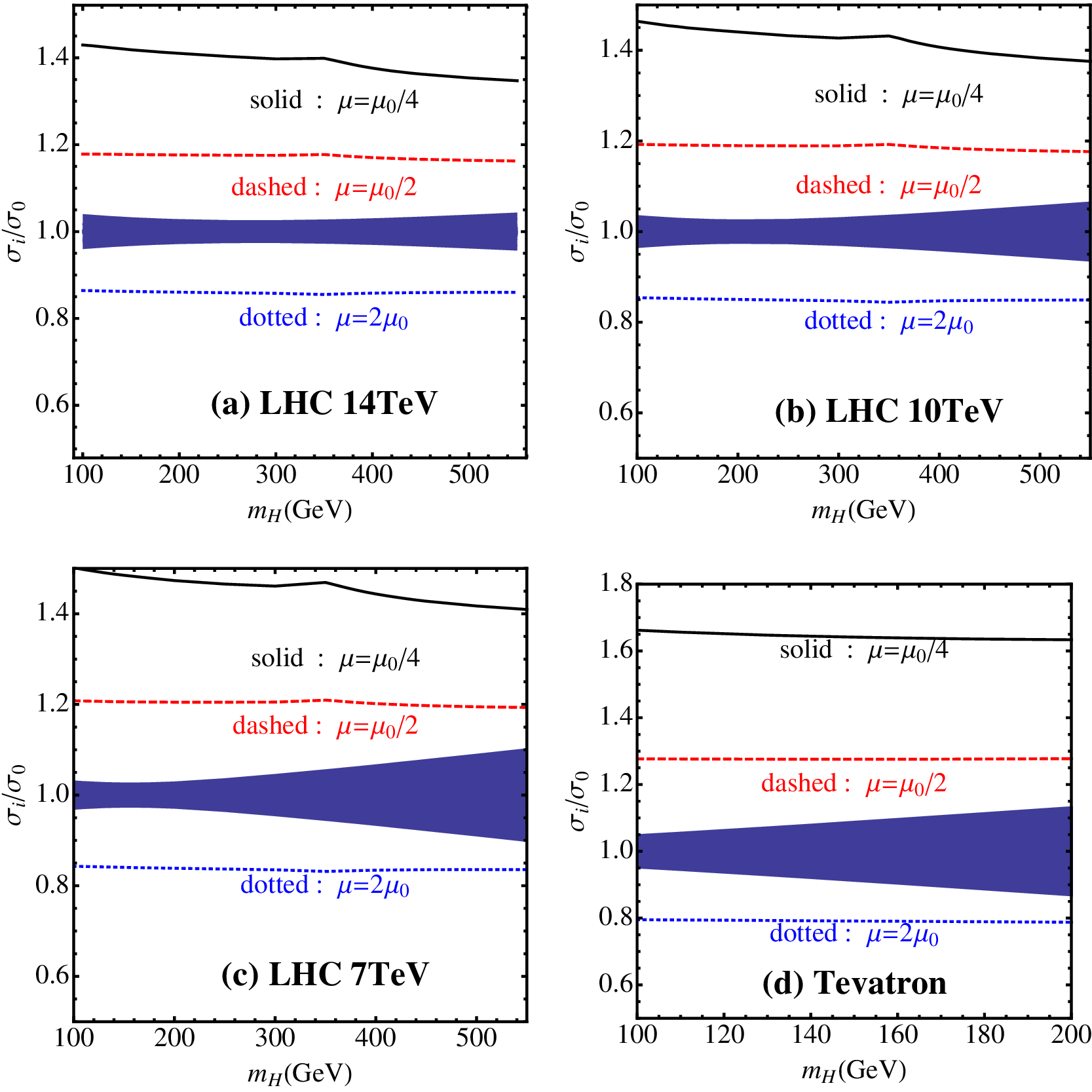}
\caption{
The PDF uncertainties of the NLO total cross section of $gg\to HX$ shown in bands  and the dependence on the renormalization scale at
the 7~TeV, 10~TeV and 14~TeV LHC and Tevatron using
the CTEQ6.6M PDFs. The renormalization and factorization scales $\mu=\mu_R=\mu_F$ are set to $\mu_0/4$, $\mu_0/2$ and $2\mu_0$, where $\mu_0 = m_H$, and the ratio is taken
with respect to the cross section evaluated with $\mu=\mu_0$.~\label{fig:scale}}
\end{figure}

In Fig.~\ref{fig:scale}, we display the uncertainties both due to the PDF uncertainties and due to the scale dependence.
The uncertainties due to the PDFs (primarily the gluon PDF), relative to the cross section with the best-fit PDF,
is shown by the bands in Fig.~\ref{fig:scale}.  The PDF error is calculated from the master formula given in Eq.~(2.5) in Ref.~\cite{Stump:2003yu},
using the 44 sets of CTEQ6.6M package.
For $100~\rm{GeV}\leq m_H \leq 600~\rm{GeV}$, the uncertainty is smaller than $5\%$ at the LHC with a c.m.~energy of both 14 TeV and 10 TeV.
In the intermediate mass region, $200\,\rm{GeV}\lesssim m_H \lesssim 300\, \rm{GeV}$, where the gluon PDF is more constrained, the uncertainty is
reduced to about $2\%\sim 3\%$
Setting $x_1 \approx x_2=x$ where $x_{1,2}$ is the momentum fraction of the incoming parton, and using $\hat{s}=x_1 x_2 s$,
we obtain $\left<x\right>\approx m_H/\sqrt{s}$.  From this we see that the minimum in the PDF uncertainty in both Fig.~\ref{fig:scale}(a)~and~(b)
occurs around $\left<x\right>\sim 0.022$.
On the contrary, the Higgs boson production cross section at the Tevatron suffers
from much larger PDF uncertainties. For example, the uncertainty increases from $5\%$ to $14\%$
for the mass of Higgs bosom mass $100\, \rm{GeV}\leq m_H \leq 200\, \rm{GeV}$.

We also display the uncertainties in the cross section calculation at NLO due to the renormalization scale ($\mu_R$) and
factorization scale ($\mu_F$) dependence in Fig.~\ref{fig:scale}.  These uncertainties can be considered as an estimate of the
size of the unknown higher order corrections.  For this study, we have set $\mu=\mu_R=\mu_F$ and vary it around the central
value of $\mu_0=m_H$.   Typically, a factor of 2 is used to estimate the size of the higher order corrections, so we have
displayed curves with $\mu=2\mu_0$ and $\mu=\mu_0/2$.  In addition, since the NNLO QCD corrections prefer
a scale of $\mu=\mu_0/4$~\cite{Anastasiou:2002yz}, we also display a curve with that value.  In Fig.~\ref{fig:scale} we plot the ratio $\sigma(\mu_i)/\sigma(\mu_0)$
as a function of $m_H$ both at the LHC and at the Tevatron.
The cross sections vary between about $ -15 \% $ for $\mu = 2\mu_0$ and
$+20\%$ for $\mu = \mu_0/2$ at the LHC, and can reach about $+40\%$ when using $\mu = \mu_0/4$.
At the Tevatron, the scale dependences are even larger.
We note that the scale dependence at both colliders is insensitive to $m_H$ and
dominates over the PDF uncertainties.

\begin{figure}
\includegraphics[scale=0.5]{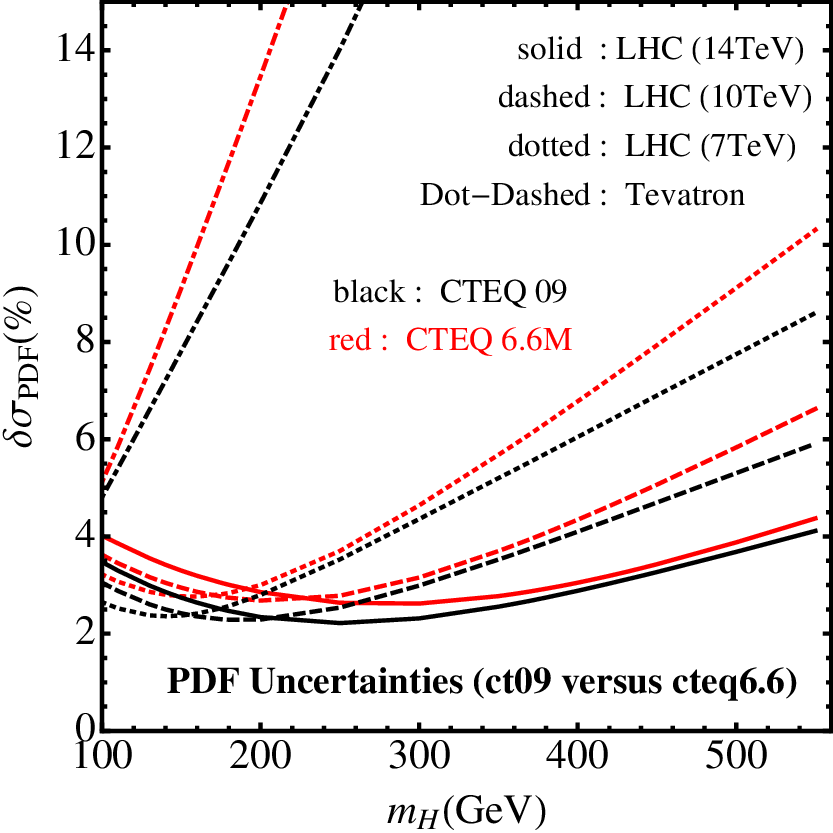}
\caption{The uncertainties of the NLO total cross section of $gg\to HX$ due to the PDFs.
The red curves are for CTEQ6.6M and the black curves are for CT09.~\label{fig:09vs66}}
\end{figure}
The PDF uncertainties can be improved further by using the new set of CTEQ PDFs
(named CT09~\cite{Pumplin:2009nk}),
which take into account the recent inclusive jet data at the
Tevatron~\cite{Abulencia:2007ez,Aaltonen:2008eq}.
Fig.~\ref{fig:09vs66} shows the relative error, $\delta\sigma$, in the Higgs boson production cross section
due to PDF uncertainties derived from CT09 (black)
and CTEQ6.6M (red). The uncertainties in the Higgs boson production cross section
are improved substantially using the new set of PDFs,
both at the LHC and at the Tevatron. Due to the modification of the gluon PDF, the minima
of the PDF uncertainties at the LHC are shifted to smaller values of the Higgs boson mass.

\end{document}